\documentclass[12pt]{article}
\usepackage[affil-it]{authblk}
\usepackage{amssymb,physics}
\begin{document}

\title{\bf On dual formulation of Axion solution to strong-CP problem}

\author{Otari Sakhelashvili}

\affil{Max-Planck-Institut f\"ur Physik, F\"ohringer Ring 6, 80805 M\"unchen, Germany}
\affil{Arnold Sommerfeld Center, Ludwig-Maximilians-Universit\"at, Theresienstra{\ss}e 37, 80333 M\"unchen, Germany}

\maketitle

\begin{abstract}

An exact duality between axion with arbitrary potential and antisymmetric form-field has been derived some time ago. Using this duality, the axion solution to strong-CP problem has been formulated as a gauge invariant theory of forms. In this description, the QCD axion is represented by a Kalb-Ramond field which is eaten-up by the Chern-Simons $3$-form of QCD, thereby making it massive. This ensures the CP-invariance of the vacuum. Although viewed as an effective low energy theory, this formulation accomplishes the same goal as ordinary Peccei-Quinn mechanism, due to its gauge invariance, it is protected against unwanted UV-corrections. In the previous work it has been shown that dual formulation is insensitive to UV-physics in the sense that the corrections to CP-conserving vacuum from arbitrary massive sources are strictly zero.  By going carefully through duality transformations and source-resolution, we reproduce this curious result and give some further consistency checks.  We apply similar analysis to other approaches to naturalness problems based on form fields and axions, such as cosmological relaxation of the standard model Higgs boson mass via the attractor mechanism.

\end{abstract}

\section{Introduction}

The dualities involving form-fields play important role in descriptions of physical systems. In \cite{Dvali:2005an} it was shown that there is an exact duality between a massive  pseudoscalar $a$ (axion) with arbitrary potential  $V(a)$ and a gauge-invariant theory of coupled antisymmetric $2$- and $3$-forms. Using this connection, in the above work, the strong-CP puzzle and its axion solution were reformulated as the gauge theory of forms. The $3$-form gauge field  $C$ is an organic part of QCD in form of the Chern-Simons three form.  In the absence of axions (or chiral massless quarks\footnote{In a chiral limit of a massless quark, QCD contains a built-in axion in form of $\eta'$-meson \cite{Dvali:2005an}.}), the $3$-form is massless.  Such a field has no propagating degrees of freedom. However, its ``electric" field-strength, $F \equiv dC$, can assume arbitrary constant value in the vacuum. In the absence of axion, this electric field {\bf $^*E=F$} cannot be changed neither classically nor quantum-mechanically. The different values of $E$ form  different superselection sectors.  Vacua with $E\neq 0$ are CP non-conserving.  Therefore, the  vacua with different values of $E$ represent alternative description of familiar $\theta$-vacua \cite{TV} of QCD. In this description, $E$ plays the same role in parameterizing CP-breaking, as the vacuum angle $\theta$ plays in the conventional picture. \\

 In formulation of \cite{Dvali:2005an}, the QCD axion is introduced in the theory in form of a Kalb-Ramond antisymmetric $2$-form $B$.  The ordinary  Peccei-Quinn global symmetry under the shift of axion, is replaced by gauge symmetry of $B$ and $C$ forms.  The $2$-form axion $B$ becomes a longitudinal (St\"uckelberg) component of the $3$-form $C$, forming a massive $3$-form field.  The massive $3$-form gauge field can no longer sustain a non-zero constant electric field $E$ in the vacuum, which becomes  screened. Thus, in the presence of $B$-axion, we end up with an unique CP-conserving vacuum $E=0$. In ordinary language,  this is equivalent to relaxing $\theta=0$. \\

 At the level of low energy effective theory, the dual formulation of  \cite{Dvali:2005an} accomplishes exactly the same goal  as the conventional axion \cite{axion} of Peccei-Quinn scenario  \cite{PQ}.  However, the sensitivities of the two formulations with respect to UV-physics can be drastically different.  In the language of original Peccei-Quinn formulation  \cite{PQ} the UV-corrections are essentially uncontrollable, due to the fact that an explicit breaking of global Peccei-Quinn symmetry is not forbidden  by any fundamental principle. For example, one can argue that in gravity such a breaking is triggered by high dimensional operators suppressed by Planck scale \cite{grav_breaking}.   

 In contrast,  the gauge symmetry of the dual formulation must be respected by arbitrary UV-completion of the theory \cite{Dvali:2005an}. This puts the potentially dangerous corrections under much stricter control as compared to ordinary formulation.  \\

  As shown in  \cite{Dvali:2005an}, due to its gauge invariance, the dual formulation is fully insensitive to UV-physics. Namely, it was proven that,  provided new physics contains no long-range $3$-form correlators (i.e., no massless $3$-form fields), the CP-invariance of the vacuum ($E=0$), is not affected at all.  That is,  CP-violating corrections to the vacuum,  coming from arbitrary  massive modes, are exactly zero.  The proof  is based on effective field theory methods relying on  gauge invariance and the analysis of the pole structure of the propagators brought by new physics. \\ 

  From the first look, the above sounds like a rather strong statement. However, the formulation in terms of $3$-form gauge theory, makes the story sufficiently transparent. In the present paper we shall reproduce this result by explicitly resolving  heavy sources and shall further elaborate on it. \\

  As discussed in \cite{Dvali:2005an}, the insensitivity of gauge formulation to massive physics gives useful tool for parameterizing and avoiding unwanted corrections to axion solution of strong-CP from the sources such as gravity. Namely, the UV-insensitivity criterion tells us that,  in dual formulation, the unwanted corrections can only appear if the theory contains additional massless $3$-forms that can also mix with the QCD axion. If such massless forms exist, they can  ``disrupt" axion from fully screening the QCD Cern-Simons $3$-form field.   For example, in gravity such a potential danger to axion can come uniquely from a gravitational Chern-Simons $3$-form, provided the latter can give a long-range correlator. \\

  As explained in \cite{Dvali:2005an}, even if gravity contains  such unwanted corrections, the dual gauge-invariant  formulation offers a way out. A disruption from gravitational Chern-Simons can be easily avoided if the theory contains an additional chiral symmetry that is anomalous with respect to gravity. In fact,  the role of a ``protector" for axion can be played by a chiral symmetry of a light fermion, such as neutrino \cite{Dvali:2005an, Dvali:2013cpa} \footnote{Possible phenomenological implications of this effect for neutrino masses were explored in \cite{Dvali:2016uhn}.}.  \\

 Another  application of $3$-form/axion system is Dvali-Vilenkin ``attractor" mechanism \cite{Dvali:2003br, Dvali:2004tma}.  Originally this mechanism was used for the cosmological relaxation of the mass of the standard model Higgs boson. This scenario is motivated by the hierarchy problem, an inexplicable smallness of the Higgs mass relative to Planck scale, an ultimate cutoff of the theory. The conventional approaches, such as low energy supersymmetry, or low scale quantum gravity \cite{Arkani-Hamed:1998jmv, Dvali:2007hz}, predict the existence of new physics not far from the weak scale.  The mechanism of \cite{Dvali:2003br, Dvali:2004tma} intended to offer a potential way to relax the Higgs mass (and the VEV) to acceptably low values, without the need of low energy stabilizing physics. The idea is to use a $3$-form field as a control parameter  for the Higgs mass. The   vacua with different values of the Higgs mass are then actualized  due to the change of the electric field $E$ by  its sources.  Their role can be played by  branes or by solitons (domain walls) of heavy axion-like fields. The key ingredient of the mechanism is that the step of the variation of $E$ diminishes towards a certain critical value  $E_*$. This results into a divergent density of vacuum states  with values of $E$ arbitrarily close to $E_*$.  Due to this, the vacuum  $E=E_*$ acts as an attractor  point of the cosmological evolution. 
  
 The $3$-form attractor mechanism has also been applied to the strong-CP problem \cite{Dvali:2005zk}. In this setup, the role of $C$ is played by the QCD Chern-Simons and the attractor point is at the vanishing electric field  $E_* =0$, corresponding to CP-conserving vacuum. The idea is that the universe can be driven to it by cosmic evolution without the need of the axion field.

 In order to represent a legitimate solution of the  naturalness problems, it is important for the attractor mechanism not to be UV-sensitive.  In particular, the attractor point should not be sensitive to the details of short distance physics such as the inner structure of the brane.  For this, it is important to understand under what circumstances the brane structure becomes important.  In the present paper we shall also address such issues. \\

 The paper is orginized as follows. After carefully reviewing the  duality between $3$-forms and axion and a dual  formulation of axion given in \cite{Dvali:2005an}, we study resolutions of the heavy sources. We observe that screening of $E$ (equivalently $\theta$) by axion in QCD is not sensitive to the presence of the heavy sources and/or massive states. We thus confirm the findings of \cite{Dvali:2005an}. \\

 We then turn to the situations when no light axion is present in low energy theory and  $C$ is sourced exclusively by massive branes.  In such a case $C$ remains massless and can produce a long-range electric field $E$. This situation does not correspond to the case of QCD axion, but  is relevant for the attractor setups \cite{Dvali:2003br, Dvali:2004tma, Dvali:2005zk}. We compute the back reaction from the electric field $E$ to the brane in the leading order and establish the time-scale after which the inner structure of the brane is affected.  Applying this to the attractor mechanism, we shall find out that the   physics near the attractor point is insensitive to this resolution. As another sort of a back reaction, the motion of branes leads to the particle-creation.  In case of \cite{Dvali:2003br, Dvali:2004tma} the radiated quanta are Higgs bosons. This is due to the change of the Higgs boson mass triggered by the electric field $E$.  Again, the effect vanishes near the attractor point but can  be significant when the system is far from it.

\section{Duality} 

 The dualities between the form-fields have long history. In particular, it has been known for a long time that a theory of a free massless $2$-form Kalb-Ramond field  $B_{\mu\nu}$,  described by the Lagrangian,
 \begin{equation}
\mathcal{L}=\frac{1}{12}(dB)^2,
\label{LB}
\end{equation}
is dual to a free massless pseudoscalar $a$,
  \begin{equation}
\mathcal{L}= \frac12(\partial a)^2\,,
\label{LaK}
\end{equation}
where $d$ denotes an exterior derivative. The proof of duality at the level of free theory is straightforward. However, for some time, the dualities at the level of massive interacting theories were a source of controversy. It has even been suggested \cite{KLLS} that duality ceases to exist at the level of an interacting theory and that  interaction and mass terms break duality.

 This issue was clarified in \cite{Dvali:2005an} where it was proven that duality continues to hold for massive and interacting axion field with arbitrary scalar potential $V(a)$. The caveat is that the  non-derivative interaction terms of the axion dualize to a non-trivial kinetic function of a massive $3$-form field $C_{\mu\nu\alpha}$ in which the Kalb-Ramond $B_{\mu\nu}$ enters as the longitudinal  (St\"uckelberg) degree of freedom.   This duality has smooth limits both for the zero coupling and the zero mass. The number of degrees of freedom remains in tact in both limits. We shall start by reviewing this duality closely following \cite{Dvali:2005an}. 
   
   Let us consider a theory of a massive pseudo-scalar axion with arbitrary  potential 
  \begin{equation}
\mathcal{L}= \frac12(\partial a)^2-V(a)\,.
\label{La}
\end{equation}
 We wish to show that this theory is dual to the following theory of an interacting massive $3$-form $C$ with the field strength $F_{\mu\alpha\beta\gamma}= \partial_{[\mu} C_{\alpha \beta\gamma]}=\epsilon_{\mu\alpha\beta\gamma}E$,  
\begin{equation}
\mathcal{L}=\Lambda^4\mathcal{K}\left(\frac {E}{\Lambda^2}\right)+\frac12m^2 C^2 \,,
\label{LCM}
\end{equation} 
 where  $\Lambda$ and $m$ are parameters of mass dimensionality. The quantity $E$ shall be referred to as the ``electric" field. The $\mathcal{K}\left(\frac E{\Lambda^2}\right)$ is a non-derivative function of its argument $E$, which satisfies, 
 \begin{equation}
V(a) =\frac{1}{\sqrt6}m\Lambda^2 \int da \, {\rm \bf inv}\, \mathcal{K}'\left(\frac{m a}{\sqrt6 \Lambda^2}\right)\, 
\label{Va}
\end{equation}
 where, prime denotes a derivative with respect to the argument and {\bf inv} stands for an inverse function. For a simplest choice $\mathcal{K}(x)=\frac12x^2$, we get quadratic and canonically normalized Lagrangian. \\

In order to prove the above, let us first decompose the massive field $C$ in its transverse and longitudinal modes, 
  \begin{equation}
C  = C^{T}  - dB\, 
\label{CTB}
\end{equation}

 The longitudinal mode, which is the only propagating mode of the massive $3$-form field, is a Kalb-Ramond field. It also serves as the St\"uckelberg field for explicitly maintaining the following gauge redundancy, 
  \begin{equation} \label{GaugeCB}
 C^{T} \rightarrow C^{T} +  d \Omega\,,~~ B \rightarrow B + 
 \Omega \,. 
\end{equation}  
Here $\Omega$ is a gauge shift parameter, which represents an arbitrary two-form function of spacetime coordinates. \\

 Notice, (\ref{CTB}) is just a decomposition, and does not amount to any modification of the theory. The original theory (\ref{LCM}), as well as its decomposed version,
\begin{equation}
\mathcal{L}=\Lambda^4\mathcal{K}\left(\frac{E}{\Lambda^2}\right)+\frac12m^2 (C^T - dB)^2 \,,
\label{LCB}
\end{equation} 
are physically equivalent for arbitrary values of the mass, including the limit $m\rightarrow 0$.  This is obvious from the fact that they give identical physical observables, such as the interactions mediated between arbitrary external sources.\\

 Following \cite{Dvali:2005an}, we now perform dualization. As the first step, we treat  $dB \equiv X$ as a fundamental $3$-form, simultaneously  imposing the Bianchi identity ($dX=0$) as a constraint through a Lagrange multiplier $a$, 
  \begin{equation}  
\mathcal{L}=\Lambda^4\mathcal{K}\left(\frac {E}{\Lambda^2}\right)+\frac12m^2 (C^T - X)^2+ \frac{1}{\sqrt{6}}m \, a \, \epsilon_{\mu\alpha\beta\gamma} \partial_{\mu} X_{\alpha \beta\gamma}
 \,.
\end{equation} 
Integrating out $X$ through the equations of motion,  we get the following effective theory, 
 \begin{equation} \label{KFa}
\mathcal{L}=\Lambda^4\mathcal{K}\left(\frac {E}{\Lambda^2}\right)+\frac12 (\partial a)^2 -\frac{1}{\sqrt{6}}m\,a\,E.
\end{equation} 
Notice that the gauge symmetry (\ref{GaugeCB}) of Kalb-Ramond field is replaced by the global shift symmetry of the axion field by an arbitrary constant, 
 \begin{equation} \label{GlobalA}
a \rightarrow a + {\rm const} \,.
\end{equation} 
  The equation of motion for $C$ gives, 
 \begin{equation}
 \label{EqF1}
\partial_{\mu}\left(\mathcal{K}'\left(\frac{E}{\Lambda^2}\right)-\frac{m}{\sqrt6\Lambda^2}a\right)=0,
\end{equation} 
whereas the one of axion is, 
  \begin{equation}
    \label{Eqa1}
 \Box a  = \frac1{\sqrt6}m E\,.  
\end{equation}
 Solving for $E$ as function of $a$ from(\ref {EqF1}) and taking into account (\ref{Eqa1}) makes it obvious that 
 \begin{equation}
 \label{F(a)}
E(a) = \frac{\sqrt6}{m} \frac{d V(a)}{da} \,. 
 \end{equation} 
 This is the source of the relation (\ref{Va}). Thus, integrating out $E$-field, we  finally arrive to the theory (\ref{La}) with the potential  $V(a)$ determined by the function $\mathcal{K}'$ through the relation (\ref{Va}).  Thus we reproduce the result of  \cite{Dvali:2005an} that theory of an axion field $a$  (\ref{La}) with an arbitrary potential, has an exact dual in form of a gauge theory (\ref{LCB}) of $C^T$ and $B$,  where the axion potential is translated into the kinetic function $\mathcal{K}$ of the $3$-form through the relation (\ref{Va}). \\
  
  Next, using the above duality, in \cite{Dvali:2005an}  the solution of the strong-CP problem was formulated as the Higgs-like phase of the gauge theory of forms. This formulation contains no reference to global chiral symmetry and only relies on gauge redundancy.   This dual formulation is described by the Lagrangian (\ref{LCB}), where $C^T$ represents a Chern-Simons $3$-form of QCD and $B$ is a Kalb-Ramond dual of the axion $a$.  The theory  (\ref{LCB}) eliminates strong-CP violation in a gauge redundant formulation. In this formulation, the global Peccei-Quinn symmetry, which acts on axion (\ref{GlobalA}) is replaced by the gauge symmetry  (\ref{GaugeCB}). Viewed as low energy theories, the two  formulations accomplish exact same goal, but  exhibit different sensitivities towards UV-physics.  In particular, in formulation (\ref{LCB}) the vacuum of the theory exhibits zero UV-sensitivity. \\

  An example of $\mathcal{K}(x)$ function, $\mathcal{K}(x) \propto  (x  \arcsin{x} + \sqrt{1 - x^2})$, provided in \cite{Dvali:2005an}, gives a simplest prototype  potential $V(a) \propto \rm \cos(\frac{a}{f_a})$ (here $f_a$ is an axion decay constant) commonly used in axion literature.   Conventionally, this potential is obtained by instanton calculation in dilute gas approximation \cite{axion_inst}. The $3$-form language shows that the vanishing of CP-violation is  insensitive to a precise form of $V(a)$.

  In QCD, the function $\mathcal{K}$ is not known exactly, but important thing is that none of the key aspects of the axion physics, such as  the generation of the axion mass gap and elimination of $\theta$, are sensitive to its exact form.  In particular, from (\ref{F(a)}) it is clear that $E=0$ at every extremum of the axion potential, regardless the form of the function $\mathcal{K}$.  This constitutes  an alternative support of Vafa-Witten theorem \cite{Vafa:1984xg}  and its generalization to all extrema of $V(a)$.    
     \\

 Thus,  the above duality allows to understand the generation of axion mass in QCD as the $3$-form Higgs effect. Notice, this  description is exact \cite{Dvali:2005an}, as the entire information about the axion mass and its potential is contained in non-derivative function $\mathcal{K}$. None of the high derivative contributions affect either the mass or the potential, since  such contributions vanish in zero momentum limit.  Thus, as long as no additional massless $3$-form fields are added to QCD, the generation of the axion mass with simultaneous elimination of the  $\theta$-vacua, can be understood in the language of the Lagrangian (\ref{LCB}) (equivalently (\ref{LCM})) or its dual (\ref{La}). 

 This language also makes it  transparent why the existence of an additional potential $\tilde{V}(a)$ (not generated via a QCD $3$-form) un-Higgses the $3$-form and re-generates  $\theta$-vacua.  Indeed, in a theory with an additional potential  
 \begin{equation}
\mathcal{L}=\Lambda^4\mathcal{K}\left(\frac{E}{\Lambda^2}\right)+\frac12 (\partial a)^2 -\frac1{\sqrt6}m\,a\,E - \tilde{V}(a)\,.
\label{can_nor}
\end{equation} 
 the $3$-form remains massless. The way to understand this in the $3$-form Higgs language is to notice that the additional potential can be traded for an additional $3$-form \cite{Dvali:2005an}. It  then becomes clear that only one superposition of the two $3$-forms can become massive by eating up a single scalar field. This is the reason why any  external explicit breaking of the axion shift symmetry (PQ symmetry) in QCD brings back the  observable CP-violation.

 \section{UV-in-sensitivity of Dual Axion Mechanism} 
 
  We now wish to discuss the point of \cite{Dvali:2005an} that the axion solution, when formulated as $3$-form gauge theory (\ref{LCB}), is insensitive towards  UV  physics. That is, the generation of the mass gap cannot be affected by the presence of the heavy fields at some mass scale $M_f$.  In particular, coupling the QCD $3$-form  to arbitrary massive branes,  cannot regenerate a non-zero CP violation.  A curious thing about this statement is that it is meant to be exact.

 Naively, one would think that the introduction of massive states  at some high scale $M_f$, can affect the low energy  physical observables by small but non-zero amount. The corrections, while possibly exponentially small or suppressed by powers of the  scale $M_f$,  are usually non-zero. In case of the  gauge formulation (\ref{LCM}) of the QCD axion, this would imply that coupling the QCD  Cern-Simons term to heavy banes would perturb the observable $\theta$-parameter by small amount and trigger the CP-violation. \\
 
 However, this is not what is happening. The CP-violating order parameter $E$ remains exactly zero, as long as new physics does not come with new massless poles in $3$-form correlators.  The proof, which can be found in \cite{Dvali:2005an}, is based on integration out of the heavy physics and writing down the effective propagator for the $3$-form $C$.  We shall not repeat it here. However, we shall do a cross check in the example below.  
 \\
 
 We can test the above claim by using the explicit resolution of the brane, analogous to the one in \cite{Dvali:2004tma}.  For this, let us assume the $C$ form in the   Lagrangian (\ref{KFa}) to stand for a QCD  Chern-Simons  field.  Let us  now add its couplings to a heavy brane coming from some UV-physics.  This heavy brane can be resolved  in form of a soliton of a heavy axion $b$ with the effective  potential $V(b)$.  The theory now becomes, 
 
 \begin{equation}
 	\mathcal{L}=-\frac{1}{48}F^2- \frac{1}{2\pi}\partial_\mu(q a 
 	+ q_b b)
 	\epsilon_{\mu\alpha\beta\gamma}  C_{\alpha \beta\gamma}+\frac{1}{2}f_a^2 (\partial a)^2 \, 
 	+ \frac{1}{2}f_b^2 (\partial b)^2-V_b(b)\,.
 \label{CAB}
 \end{equation}
 Here $f_a$ and $f_b$ are decay constants of $a$ and $b$ axions which are not canonically normalized. $\mathcal{K}(x)=\frac12x^2$ and we also parameterize the mixing between $3$-form and axions with  constants $q$ and $q_b$ with mass dimension-2.  The connection between the parameters of (\ref{CAB}) and (\ref{KFa}) is straightforward.
 
 Solving the equations of motion shows that, despite the presence of the heavy field $b$, in the vacuum state the electric field $E$ of $C$ remains exactly zero. This is immediately apparent from the equation,  
 \begin{equation}
 	\label{Eqa}
 	f_a^2 \Box a  = \frac{q}{2\pi} E\,,  
 \end{equation}
 which shows that no vacuum can be reached for any non-zero value of $E$. This zero sensitivity of $E=0$ vacuum towards heavy physics (in this case heavy axion $b$)  is again very transparent  in the Language of $3$-form Higgs effect. Using the exact duality, we can rewrite the potential for axion $b$ as a Higgs effect with respect to a second $3$-form  $C^b$. The Lagrangian then becomes, 
 \begin{equation} \label{abCCb}
 	\begin{split}
 		\mathcal{L}=-\frac{1}{48}F^2- \frac{1}{2\pi}\partial_\mu(q a + q_b b)\epsilon_{\mu\alpha\beta\gamma}  C_{\alpha \beta\gamma}+\frac{1}{2}f_a^2 (\partial a)^2 \,+\\
 		+ \frac{1}{2}f_b^2 (\partial b)^2 - \frac{q_{bb}}{2\pi}\partial_\mu b \epsilon_{\mu\alpha\beta\gamma} C^b_{\alpha \beta\gamma} +\Lambda^4\mathcal{K}_b\left(\frac{E^b}{\Lambda^2} \right)\,  
 	\end{split}
 \end{equation}

  where $\mathcal{K}_b$ satisfies the same condition (\ref{Va}) with respect to $V_b(b)$. This  theory describes a Higgs effect with two gauge three-forms ($C$ and $C_b$) and two axions ($a$ and $b$). As a result, no massless degree of freedom is remaining in the theory and there exist no electric field in the vacuum with respect to any $3$-form.  Thus, all CP-violating order parameters are exactly zero.  
 
  Situation would change drastically, if the new physics, coupled to ordinary axion, would come in form of a $3$-form without its own axion partner, 
 \begin{equation}
 	\label{aCCb}
 	\mathcal{L}=-\frac{1}{48}F^2- \frac{1}{2\pi}\partial_\mu a \epsilon_{\mu\alpha\beta\gamma} (q C_{\alpha \beta\gamma}+q_b C^b_{\alpha \beta\gamma})  -\frac{1}{48}F_b^2+\frac{1}{2}f_a^2 (\partial a)^2 \,  
 \end{equation}
 This is an example considered in \cite{Dvali:2005an} as violating the criterion of absence of massless $3$-forms.  In such a case,  the new physics will be represented solely by $C_b$  which has no axion partner.  Correspondingly, there is  a massless pole that new physics brings in.   Integrating out $C_b$ and $a$, we recover (up to notations) the  equation (35) of \cite{Dvali:2005an},
 
 \begin{equation} \label{EffC1}
 	\mathcal{L}=-\frac{1}{48}F^2- \frac{1}{48}\frac{q^2}{q_b^2}F_{\mu\alpha \beta\gamma} \frac{m^2}{\Box +m^2} F_{\mu \alpha\beta\gamma} \,.
 \end{equation}
 
  Where $m^2=\frac{q_b^2}{4\pi^2f_a^2}$.  Since in zero momentum limit ($\Box \rightarrow 0$) the only effect of the second term is that it corrects kinetic term normalization, the theory contains a long-range correletor in form of a massless $3$-form field.  Correspondingly, there exists a vacuum solution with a constant  $E$, which violates CP-symmetry. 
 
 At the same time, if we perform analogous integration in (\ref{abCCb}), assuming $\mathcal{K}_b(x)=\frac12x^2$, the effective theory for  $C$ will be, 
 
 \begin{equation} \label{EffC2}
 	\mathcal{L}= \frac12 C_{\alpha \beta\gamma} \left(\Box +m^2_a+\left(\frac{q_b}{q_{bb}}\right)^2\frac{m^2_b\Box}{\Box+m^2_b}\right) \Pi_{\alpha\mu}C_{\mu \beta\gamma} \,,
 \end{equation}
  Where $\Pi_{\alpha \mu} = \eta_{\alpha \mu} - \frac{\partial_\alpha \partial_\mu}{\Box}$ is a transverse projector, $m_a^2=\frac{q^2}{4\pi^2 f_a^2}$ and $m_b^2=\frac{q_{bb}^2}{4\pi^2 f_b^2}$. This theory describes a theory with mass gap in which the longitudinal mode has been integrated out. There is no constant electric field in the vacuum and correspondingly no CP-violation. This confirms the statement of \cite{Dvali:2005an} about the insensitivity of the dual formulation of axion with respect to  heavy physics. 
 
 \section{More on Resolving Sources}

 Our next steps are the following.  First we consider a three form sourced by a brane in the zero width  approximation.  We then resolve the brane in form of a soliton as this was done in \cite{Dvali:2004tma}.  As shown there, we obtain that the topological charge of the soliton in low energy theory acts as an electric (Noether) charge for  the $3$-form field. We then take into account the back reaction that  branes experience from the $3$-form dynamics and we ask how sensitive the resolution is to this back reaction.
 
  For resolving the brane as a soliton of heavy axion like field $a$ we choose the Lagrangian in the following form \cite{Dvali:2004tma},
\begin{equation}
\mathcal{L}=-\frac{1}{48}F^2-\frac{q}{2\pi}\partial_\mu a \epsilon_{\mu\alpha\beta\gamma}  C_{\alpha \beta\gamma}+\frac{1}{2}f_a^2 (\partial a)^2-{V}(a),
\end{equation} 
 the field $a$ is a non-canonically normalized axion, with decay constant $f_a$ and a periodic potential ${V}(a)$,  which we choose in the form, 
 \begin{equation} 
 V(a)=V_0[1-cos(a)]\,,
 \label{Vofa}
 \end{equation}
where $V_0$ is constant  with the dimension of the energy density. The parameter $q$ is a coupling between the $3$-form and the axion. The connection between the parameters of the above and (\ref{can_nor}) is straightforward. This is a typical sine-Gordon potential, which we choose as a simple prototype, but our conclusion hold for arbitrary periodic potential. \\
  
Equation of motions for $C$ field will give
\begin{equation}
-\partial_\mu F_{\mu\alpha\beta\gamma}=J_{\alpha\beta\gamma},
\label{dFJ}
\end{equation}
Where $J$ is axionic current  
\begin{equation}\label{axionic_current_3form1}
J_{\alpha\beta\gamma}=-\frac{q}{2\pi}\epsilon_{\mu\alpha\beta\gamma}\partial_\mu a\,.
\end{equation}
 This current is trivially conserved due to its topological nature.

  For the full treatment we must solve the coupled system of  equations of motion.  We shall however proceed iteratively, by  first ignoring the back-reaction from $F$ to $a$.  The field $a$ will be replaced by the sine-Gordon soliton, 
 \begin{equation}
a(z)=4 \tan^{-1}e^{\frac{\sqrt{V_0}(z-z_0)}{f_a}}+2\pi N,
\label{SG}
\end{equation}
 which depends on a single coordinate $z$. This soliton describes a topologically non-trivial configuration in which the axion field $a$ interpolates between  the two nearest neighbouring minima of the periodic potential,   
\begin{equation}
\begin{array}{ll}
a=2\pi N ,& z=-\infty \\
a=2\pi(N+1) ,& z=\infty,
\end{array}
\end{equation} 
 where $N$ is an integer.  The parameter $z_0$ is arbitrary and represents a collective coordinate of the soliton. The configuration (\ref{SG}) solves the equation motion for $a$ in the limit $q=0$, when the back reaction from the $3$-form can be ignored. 

 Substituting the solution  (\ref{SG})  into the topological current we get an external source for $C$.    This approximation amounts to a limit in which we take $\frac{\sqrt{2V_0}}{f_a}\rightarrow\infty$ and keeping $q$ small. In this approximation, the soliton field acts as an external source for the $3$-form, while experiencing no back reaction from it. In fact in this limit, we have 
\begin{equation}\label{approx_domain_wall1}
 a'(z)=2\pi\delta(z-z_0) \,, 
\end{equation}
 where $'$ denotes a derivative with respect to $z$ and the sine-Gordon soliton becomes effectively a delta-function source.

 The equation of motion (\ref{dFJ}) and are gauge redundant and require gauge fixing. We fix the Coulomb gauge. Then using the substitution of $C_{\alpha\beta\gamma}=\epsilon_{\mu\alpha\beta\gamma}C_\mu$ and taking into account  (\ref{approx_domain_wall1}),  the above equation reduces to
\begin{equation}
\epsilon_{z\alpha\beta\gamma} C_z''=J_{\alpha\beta\gamma}
=  -q \delta(z-z_0) \epsilon_{z\alpha\beta\gamma}.
\end{equation}
This equation is identical to an equation of an electrostatic field produced by a static point charge in one space dimension with coordinate $z$.

 The solution for $C_z$ is, 
\begin{equation}
C_z=-\frac12 q |z-z_0|-4E_0 (z-z_0)+c_2 \,,
\end{equation}
where $E_0$ and $c_2$ are integration constants.   The electric field corresponding to this solution has the form, 
\begin{equation}\label{el_field_sol}
E=\frac18q \, sign(z-z_0)+E_0.
\end{equation}
 Thus, the topological charge of the soliton effectively  acts as a electric Noether charge for the $3$-form gauge field.

\subsection{Back Reaction} 

 We now wish to take into account the leading back reaction in $q$ on the soliton from the electric field.  In particular, the back-reaction is expected to take place because of the difference between the electric fields on the two sides of the kink. This creates an energy difference which shall accelerate the kink. In other words, the kink carries a charge under the $3$-form electric field and is pushed towards infinity.  \\ 

This effect can be taken into account by considering an effective theory for a time-dependent fluctuation of the collective coordinate,  
\begin{equation}
z_0\rightarrow z_0+ \delta z_0(t)\,.
\end{equation}
Inserting this anzats back into the Lagrangian together with three-form solution, we get an effective Lagrangian for the fluctuation
\begin{equation}
\mathcal{L}=-\frac{1}{48}F^2-\frac{q}{2\pi}aE+\frac12 f_a^2 a'^2 \dot{\delta z_0}^2- \frac{1}{2}f_a^2 a'^2-V(a).
\end{equation}
In the above Lagrangian, first term is constant and the last two terms give a total derivative, since the fluctuation of Kink does not change its topological properties.  We are therefore left with a simpler Lagrangian density, 
\begin{equation}
L=\int dz \mathcal{L}=\int dz\left(\frac12 f_a^2 a'^2 \dot{\delta z_0}^2-\frac{ q}{8\pi}a'C_z\right),
\end{equation}
Next we integrate over $z$ coordinate. In order to resolve the square of the delta-function in the first term, we explicitly take into account the profile (\ref{SG}) of the sine-Gordon soliton and integrate over it. In the second term, we can perform the integration within delta function approximation  (\ref{approx_domain_wall1}). Choosing proper normalization we get following Lagrangian,
\begin{equation}
L=\frac12 M \dot{\delta z_0}^2+\frac{1}{f_a^2} \left(\frac18q^2 |\delta z_0|+qE_0 \delta z_0\right),
\end{equation}
this system can be understood as a point charge with mass $M=8 \frac{\sqrt{V_0}}{f_a}$ and  coordinate $\delta z_0$ in one dimensional external electric field \cite{Landau:1975pou}, 
For $\delta z_0>0$ the above system has solutions in which the soliton (domain wall) is moving accelerated 
\begin{equation}\label{kink_accelerated}
\delta z_0(t)=\frac{1}{2M f_a^2} \left(\frac18 q^2+ qE_0 \right)t^2 +vt +\delta z_0(0)\,.
\end{equation}
where $v$ is the initial velocity. Without loss of generality,  let us assume that the initial coordinate and initial velocity are zero. \\

Let us try to identify the validity time-scale of the above approximation. This is the time-scale after which the effective theory in which we treat soliton without changing internal structure breaks down.

 Most conservatively, this can be estimated by demanding that the variation of the collective coordinate should not exceed the width of the soliton. This gives a condition, 
\begin{equation}
\frac{f_a}{\sqrt{V_0}} \gtrsim |\delta z_0|,
\end{equation}
which applied to the  solution (\ref{kink_accelerated}) implies, 
\begin{equation}
\frac{f_a}{\sqrt{V_0}} \gtrsim \frac{1}{16\sqrt{V_0} f_a} \left|\frac18 q^2+ qE_0 \right|t^2\,.
\end{equation}
 This leads us to the following upper bound on the validity time-scale, 
\begin{equation}\label{time_break}
t\lesssim \frac{4f_a}{| q|\sqrt{\left|\frac18+\frac{E_0}{q}\right|}} 
\equiv t_*\,. 
\end{equation}
After this time, the internal structure of the soliton is affected by the action of the electric field. Of course,  for a long distance observer the description of the soliton as of a point particle 
accelerated by a constant electric field, is still valid. The full relativistic solution for such a particle is well-known  \cite{Landau:1975pou}.

\section{Implications for Cosmic Attractor}

 We shall now apply the above results to the attractor mechanism \cite{Dvali:2003br, Dvali:2004tma}. 
 The idea of the attractor is that the charge of the brane $q$ with respect to a massless $3$-form $C$, through a certain chain of influences, effectively depends on the electric field $q(E)$.   An important point is that for certain critical value $E_*$, it vanishes.  That is $q(E_*) =0$. Since the change of the field $E$ among the neighbouring vacua is given by the charge, 
\begin{equation}
 \label{Step}
 \Delta E \propto q(E) \,,
\end{equation} 
the step vanishes when we approach $E_*$. That is, the number of distinct vacua required to traverse through for reaching $E_*$, is infinite. In other words, there exist infinitely many vacua with values of the electric field that are arbitrarily close to the attractor value $E_*$. \\
 
One of the requirements to the attractor mechanism is that it should not  be sensitive to UV physics. That is, the singularity in number of vacua at the attractor point   $E_*$, should be trusted without knowing the internal structure of the  brane.  We can now explicitly verify this by evaluating the scaling of the critical time $t_*$ near the attractor point. Since at the attractor point the electric field is finite  $E_*$, whereas the brane change $q(E_*)$ vanishes, it is clear from  (\ref{time_break}) that   $t_* \rightarrow \infty$.   That is, closer we are to the attractor vacuum, longer is takes for the thin-brain approximation  to break down.

\subsection{Effect of Particle Creation}

An important ingredient of the attractor solution to the hierarchy problem \cite{Dvali:2003br, Dvali:2004tma}, is that the value of the Higgs mass and the VEV is determined by the $3$-form electric field.  In this way, the Higgs mass changes from vacuum to vacuum. 

Thus, in the brane background the Higgs mass is effectively dependent on the space-time coordinates. The part of the Lagrangian in which this information is encoded has the following form
\begin{equation}
\mathcal{L}=\frac12 (\partial h)^2 -\frac12 h^2\left(m_h^2+\frac{F^2}{48M_f^2}\right),
\end{equation}
where $h$ is a Higgs field, $m_h$ is its ``bare"  mass and $M_f$ is some fundamental scale, which will be assumed to be large.

Taking in the account solution for electric field (\ref{el_field_sol}), we get effectively a mass term for Higgs which depends on a position on $z$-axis.  Basically, since the electric field changes across the brane, so does the Higgs mass. Correspondingly, we can define two  different masses $M_-$ and $M_+$,
\begin{equation}
M_\pm^2=m_h^2-\frac{1}{128}\frac{q^2}{M_f^2}-\frac{E_0^2}{2M_f^2} \pm \frac18 \frac{E_0 q}{M_f^2} \,,
\end{equation}
on the left and right sides of the brane respectively. 

Without loss of generality,  we assume that the constant part of the electric field is positive.  For the opposite case we could just change the labeling. Let us assume that for some initial time $t=0$, on both sides of the wall the Higgs field is in corresponding vacuum states. That is, no Higgs particles are excited. 
 
Now, since the wall is moving accelerated (\ref{kink_accelerated}), the electric field changes in time, and correspondingly changes the vacuum of the Higgs field. This leads to a particle creation.

In order to compute the rate of particle-creation, let us start with a region in the vacuum which corresponds to $M_+$.  After the brane passes-by this region,  the mass decreases to $M_-$. 
  
In this case the number operator of $M_-$-mass particles is defined in the following way
\begin{equation}
N=\int \frac{d^3p}{(2\pi)^3}\frac{1}{2\omega_-}a^+_-(\vec{p})a_-(\vec{p})\,,
\end{equation}
where $\omega_\pm^2=p^2+M_\pm^2$, and the subscript on the ladder operators has the same meaning. Performing a Bogoliubov transformation (\ref{num_bog}) and averaging in the vacuum $\ket{0_+}$ (which is the vacuum in this case), we get following expression,
\begin{equation}
N=SL\frac14 \int d^3 p \frac{\omega_+}{\omega_-}\left(\frac{\omega_-}{\omega_+}-1\right)^2 \,,
\end{equation}
where $S$ is the surface area of the brane and $L$ is the length traveled by it in $z$ direction. Taking into account that 
\begin{equation}
\omega_+^2-\omega_-^2=\frac1{16} \frac{E_0q}{M_f^2}\,,
\end{equation}
for  $M_f^2 \gg \frac{E_0q}{M_-^2}$ we get
\begin{equation}
\omega_+\approx \omega_-+\frac1{32} \frac{E_0q}{M_f^2\omega_-}\,.
\end{equation}
At the lowest order, the produced particle number per unit surface is given by
\begin{equation}
\frac{N}{S}=L\frac1{4096}\frac{E_0^2q^2}{M_f^4} \int d^3 p \frac{1}{\omega_-^4}\,.
\end{equation}
After integration we obtain, 
\begin{equation}
\frac{N}{S}=L\frac{\pi^2}{4096}\frac{E_0^2q^2}{M_f^4M_-}\,.
\end{equation}
Substituting the distance traveled by the domain wall in time $t$,  we get the final result
\begin{equation}
\frac{N}{S}=\left(\frac{1}{16\sqrt{V_0} f_a} \left(\frac18 q^2+ qE_0 \right)t^2 +vt \right)\frac{\pi^2}{4096}\frac{E_0^2q^2}{M_f^4M_-}.
\end{equation}
The above expression gives us the amount of Higgs particles created by a passing-by domain wall.

However, we should remember that  we can trust the above analysis until the time (\ref{time_break}). During this time, we get following amount of particles created per unit surface
\begin{equation}
\frac{N}{S}=\frac{f_a}{\sqrt{V_0}}\frac{\pi^2}{4096}\frac{E_0^2q^2}{M_f^4M_-}\,.
\end{equation}
 Far from the attractor,  where $q$ is large, the effect of particle creation can be  significant and affect the dynamics of the system. However, not surprisingly, since the effect is proportional to $q^2$, it is vanishingly small near the attractor point.  Again, the attractor behave is largely unaffected by the  high energy effects. \\
 
 \section{Conclusions} 
 
 The dual formulation of the axion solution of the strong-CP problem in form of a gauge theory of a $3$-form \cite{Dvali:2005an}, by the power of  gauge invariance, gives a possibility of controlling the  unwanted UV-corrections.  In particular, it has been shown that corrections from arbitrary massive physics to CP-invariance of the vacuum are exactly zero.  In the present paper we reproduced this result  and performed consistency checks on some explicit examples of heavy physics. In accordance with the proof given in \cite{Dvali:2005an}, we find that the only possibility of destabilizing the CP-invariant vacuum is through the appearance of additional massless $3$-form fields coupled to axion.  In their absence, the QCD vacuum remains exactly  CP-conserving.
 \footnote{We note that, perhaps, an alternative physical way of understanding this UV-insensitivity is that the generation of axion mass from $3$-form Higgs effect can be understood in purely topological terms, as discussed in \cite{Dvali:2005ws}.} \\
  
We also gave a resolution of brane sources coupled to massless $3$-form fields and estimated a back reaction from the $3$ electric field to leading order.  Such setup  has been applied to the solutions of the hierarchy \cite{Dvali:2003br, Dvali:2004tma} and strong-CP \cite{Dvali:2005zk} problems via the attractor mechanism.  We observe that the existence of the attractor point is not sensitive to the brane resolution or to back reaction.  We also estimated particle-creation in the background of time dependent $3$-form electric field.

\section*{Acknowledgements}
 
We would like to thank Gia Dvali for many illuminating discussions. We also thank Lasha Berezhiani and Georgios Karananas for valuable discussions. We would like to mention that general discussions about the strong CP problem with Cesar Gomez and Goran Senjanovic were important.

\appendix

\section{Bogolyubov Transformations}
 In the above section, we have a scalar field the mass of which effectively depends on time. This means that this field does not have a fixed vacuum. Using Bogolyubov transformations,  we can properly account for the change of the vacuum. Let us consider a free scalar field $\phi$, in Schr\"odinger picture
\begin{equation}
\phi(\vec{x})=\int \frac{d^3 p}{(2\pi)^3}\frac{1}{2\omega(p,t)}(a_t(\vec{p})e^{i\vec{p}\vec{x}}+a^+_t(-\vec{p})e^{i\vec{p}\vec{x}})\,,
\end{equation} 
  where $\omega(\vec{p},t)$ is the time depended frequency. Therefore, ladder operators depend on time as well (despite being  described in Schr\"odinger picture). The corresponding canonical momenta have the following form
\begin{equation}
\pi(\vec{x})=\int \frac{d^3 p}{(2\pi)^3}\frac{-i}{2}(a_t(\vec{p})e^{i\vec{p}\vec{x}}-a^+_t(-\vec{p})e^{i\vec{p}\vec{x}})\,.
\end{equation}
  The important point here is that the field and the canonical momenta do not depend on time explicitly.  Thus,  for any given time we can write
\begin{eqnarray}
\frac{1}{2\omega(p,t)}\left(a(\vec{p})_t+a^+(-\vec{p})_t\right)=\int d^3 x e^{-i\vec{p}\vec{x}}\phi(\vec{x})\nonumber\\
\frac{-i}{2}\left(a(\vec{p})_t-a^+(-\vec{p})_t\right)=\int d^3 x e^{-i\vec{p}\vec{x}}\pi(\vec{x})\,.
\end{eqnarray}
 From this equation, immediately follows the following transformation
\begin{eqnarray}
a(\vec{p})=\frac{1}{2}\left(\frac{\omega}{\omega_0}+1\right)a_0(\vec{p})+\frac{1}{2}\left(\frac{\omega}{\omega_0}-1\right)a_0^+(-\vec{p})\nonumber\\
a(-\vec{p})^+=\frac{1}{2}\left(\frac{\omega}{\omega_0}+1\right)a_0(-\vec{p})^++\frac{1}{2}\left(\frac{\omega}{\omega_0}-1\right)a_0(\vec{p}),
\end{eqnarray}
where subscript $0$ means that the quantities are evaluated at $t_0=0$. The quantities without subscript are evaluated at time $t$ and all frequencies are evaluated for momentum $p$.  We can compute different quantities after we get connection between these operators. One example is the number operator, which at  the time $t$ has following form
\begin{equation}
N=\int \frac{d^3p}{(2\pi)^2}\frac{1}{2\omega}a^+(\vec{p})a(\vec{p}).
\end{equation}
Rewriting it in the terms of operators in time $t_0=0$, we get
\begin{equation}\label{num_bog}
\begin{split}
N=\frac14 \int \frac{d^3p}{(2\pi)^2}\frac{1}{2\omega}\left[\left(\frac{\omega}{\omega_0}+1\right)a_0(\vec{p})^++\left(\frac{\omega}{\omega_0}-1\right)a_0(-\vec{p})\right]\\
\left[\left(\frac{\omega}{\omega_0}-1\right)a_0(-\vec{p})^++\left(\frac{\omega}{\omega_0}+1\right)a_0(\vec{p})\right]\,,
\end{split}
\end{equation}
which shows that in vacuum, corresponding $t_0=0$ after certain time particles will be created.


\begin{thebibliography}{99}
	

\bibitem{Dvali:2005an}
G.~Dvali,
``Three-form gauging of axion symmetries and gravity,''
[arXiv:hep-th/0507215 [hep-th]].

\bibitem{TV}
C.G. Callan, R.F. Dashen and D.J. Gross, Phys. Lett. B63 (1976) 334;
R. Jackiw and C. Rebbi, Phys. Rev. Lett. 37 (1976) 172.


\bibitem{axion} S. Weinberg, Phys. Rev. Lett. 40 (1978) 223;

F. Wilczek, Phys. Rev. Lett. 40 (1978) 279

\bibitem{PQ}
R.D. Peccei and H.R. Quinn, Phys. Rev. Lett. 38 (1977) 1440; Phys. Rev. D16
(1977) 1791.

\bibitem{grav_breaking}
M.~Kamionkowski and J.~March-Russell,
Phys. Lett. B \textbf{282} (1992), 137-141

R.~Holman, S.~D.~H.~Hsu, T.~W.~Kephart, E.~W.~Kolb, R.~Watkins and L.~M.~Widrow,
Phys. Lett. B \textbf{282} (1992), 132-136

S.~M.~Barr and D.~Seckel,
Phys. Rev. D \textbf{46} (1992), 539-549


\bibitem{Dvali:2013cpa}
G.~Dvali, S.~Folkerts and A.~Franca,
``How neutrino protects the axion,''
Phys. Rev. D \textbf{89}, no.10, 105025 (2014)
doi:10.1103/PhysRevD.89.105025
[arXiv:1312.7273 [hep-th]].
 


\bibitem{Dvali:2016uhn}
G.~Dvali and L.~Funcke,
``Small neutrino masses from gravitational \ensuremath{\theta}-term,''
Phys. Rev. D \textbf{93}, no.11, 113002 (2016)
doi:10.1103/PhysRevD.93.113002
[arXiv:1602.03191 [hep-ph]].

\bibitem{Dvali:2003br}
G.~Dvali and A.~Vilenkin,
``Cosmic attractors and gauge hierarchy,''
Phys. Rev. D \textbf{70}, 063501 (2004)
doi:10.1103/PhysRevD.70.063501
[arXiv:hep-th/0304043 [hep-th]].


\bibitem{Dvali:2004tma}
G.~Dvali,
``Large hierarchies from attractor vacua,''
Phys. Rev. D \textbf{74}, 025018 (2006)
doi:10.1103/PhysRevD.74.025018
[arXiv:hep-th/0410286 [hep-th]].

\bibitem{Arkani-Hamed:1998jmv}
N.~Arkani-Hamed, S.~Dimopoulos and G.~R.~Dvali,
``The Hierarchy problem and new dimensions at a millimeter,''
Phys. Lett. B \textbf{429}, 263-272 (1998)
doi:10.1016/S0370-2693(98)00466-3
[arXiv:hep-ph/9803315 [hep-ph]].


\bibitem{Dvali:2007hz}
G.~Dvali,
``Black Holes and Large N Species Solution to the Hierarchy Problem,''
Fortsch. Phys. \textbf{58}, 528-536 (2010)
doi:10.1002/prop.201000009
[arXiv:0706.2050 [hep-th]].

\bibitem{Dvali:2005zk}
G.~Dvali,
``A Vacuum accumulation solution to the strong CP problem,''
Phys. Rev. D \textbf{74}, 025019 (2006)
doi:10.1103/PhysRevD.74.025019
[arXiv:hep-th/0510053 [hep-th]].


\bibitem{KLLS}
R. Kallosh, A. Linde, D. Linde and L. Susskind, Phys. Rev. D52 (1995) 912,
hep-th/9502069.


\bibitem{axion_inst}
C.~G.~Callan, Jr., R.~F.~Dashen and D.~J.~Gross,
Phys. Rev. D \textbf{17} (1978), 2717
doi:10.1103/PhysRevD.17.2717

S.~Coleman,
``Aspects of Symmetry: Selected Erice Lectures,''
doi:10.1017/CBO9780511565045


\bibitem{Vafa:1984xg}
C.~Vafa and E.~Witten,
``Parity Conservation in QCD,''
Phys. Rev. Lett. \textbf{53}, 535 (1984)
doi:10.1103/PhysRevLett.53.535


\bibitem{Dvali:2005ws}
G.~Dvali, R.~Jackiw and S.~Y.~Pi,
``Topological mass generation in four dimensions,''
Phys. Rev. Lett. \textbf{96}, 081602 (2006)
doi:10.1103/PhysRevLett.96.081602
[arXiv:hep-th/0511175 [hep-th]].


\bibitem{Landau:1975pou}
L.~D.~Landau and E.~M.~Lifschits,
``The Classical Theory of Fields''



\end{thebibliography}
\end{document}